# Unification and Extension of Classic Information Principles

Jianfeng Xu[1*]

[1] Information Technology Service Center of People's Court,
Beijing 100745, China.
Corresponding author(s). E-mail(s): xjfcetc@163.com;

**Abstract** To formulate an universal framework of information theory is beneficial. This study proves that the sextuple model of the objective information theory (OIT) is a sufficient and necessary condition for discussing information with four basic postulations. It is demonstrated for each metric defined in the OIT, there is a corresponding example in classical information theories or commonly used principles. Furthermore, atomic information is defined as the indivisible elementary information and the volume additivity is proven for combinations of atomic information. Consequently, the information volume that a single quantum carrier can carry is derived and a theorem relating information volume to mass, energy, and time is proved. All these efforts illustrate that the OIT is a novel information theory that can unify a variety of classical information principles and even accurately reveal the quantitative relationship between information, matter and energy.

## 1 Introduction

The rapid development of information science and technology has brought mankind into the information age. Classical principles, such as Shannon information theory[1][2], the radar equation[3][4], Nyquist sampling theorem[5][6], Metcalfe's law[7][8] and Kalman filtering[9][10], have played substantial roles in the design and development of information systems in their respective fields. However, to date, our understanding of what information is can vary widely[11][12][13][14][15][16][17][18][19][20][21][22][23]. Various classical information science theories lack internal close links and have not yet formed a unified theoretical system to effectively guide the practice of complex information system engineering. With the information system permeating into every corner of human society, it is increasingly difficult for people to penetrate and grasp the information system, especially systems composed of many information systems, called systems of systems (SoSs). It is becoming increasingly difficult for people to use a set of methods based on the unified information concept and model, similar to the traditional dynamics used to guide large-scale mechanical engineering, to guide and standardize the design, development, application and evaluation of large-scale information SoSs engineering. First, there is an urgent need to reach a basic consensus on the essence of information. Only by reaching a general consensus on the concept, structure and nature of information will it be possible to comprehensively study and make full use of this most important resource element in the information age. Second, it is urgent to form a relatively complete information measurement system. Only by forming a measurement system with a clear definition and rich types can we provide a basic reference for the study of information mechanisms and the analysis of the basic efficacy of information systems. Third, a theoretical system of information system dynamics (ISD) must be established. Only by using abundant mathematical tools and methods can we establish the corresponding information flow configuration and measurement efficiency model for each link of information collection, transmission, processing, storage and action. Only in this way can we meet the analysis and research requirements of integrating many information systems belonging to different fields and functions into a single large-scale system.

In view of these apparent requirements, this author proposed the fundamental concept of objective information theory (OIT) [24], established a definition and model of information, analyzed the basic properties of information, and defined nine metrics of information. In the context of aerospace control systems [25], they verified the applicability and reasonableness of OIT. In [26], conditions on information restorability were introduced, and eleven types of information metrics were revised and expanded, including volume, delay, scope, granularity, variety, duration, sampling_rate, aggregation, coverage, distortion, and mismatch. Further efficacy analyses were conducted for these eleven types of metrics that an information system can potentially achieve. The structure of ISD was also established, forming a theoretical framework for the study of information systems dynamics, which would provide comprehensive index guidance and model support for the design and integration of large-scale information systems. The results have been applied and verified in the Smart Court SoSs Engineering Project of China [27].

In this paper, based on the fundamental concepts of OIT, we illustrate the specific content of information through a simple real-world example and formulate four basic postulates on information: binary attributes, existence time duration, state representation, and enabling mapping. We then prove that a sextuple model realizes these postulates as necessary and sufficient conditions. According to the definition of eleven information metrics, the relationship between the metrics and the related classical or commonly used information principles is demonstrated, which shows that the information metrics defined by OIT can also express many classical principles. Finally, the using combination and decomposition properties of information, we verify the additivity of atomic information volume and prove the information volume formula for a single quantum carrier over a time period via the Margolus–Levitin theorem. Additionally, a formula regarding the relationship between information volume and mass, energy, and time is derived via Einstein's mass–energy conversion formula. The proposed formula enables us to calculate and analyze the information volume the universe can carry to date, reveals more accurate relationships among information, matter, energy, and time, and can be applied widely.

## 2 Concept, model and definition of information

While different people may think very differently about what information is, there are typical and universally acknowledged examples of information that are collected, transmitted, processed, stored, and utilized in computers and over the internet, including various types of numerical data, texts, audio, video, and multimedia files. Wiener [28] mentioned that information is just information, not matter or energy, raising information to the same level as matter and energy. [29] emphasized that matter, energy and information are three kinds of societal resources. This author believes that matter is the original existence while energy is the existence of the capability of motion, information is the objective reflection of the content of both the objective and subjective worlds, and matter, energy, and information form the three essential elements of the objective world[24].

**2.1 Concept of information: a simple example**

To have a more intuitive understanding of this concept and model of information, we can illustrate it using a penguin picture file on a laptop computer. Figure 1 displays the information of penguins and their motions on a laptop computer screen stored in the computer. The noumena of the information consists of three penguins under blue sky and white clouds, perhaps with their unknown subjective emotions. Assuming that the picture is taken at time t0 and that the shutter speed is 1% of a second, the state of the information noumena occurs in the time interval [t0; t0+0.01] seconds. The state set of the noumena includes, on the left, three the positions of three penguins under blue sky and white clouds in the same time interval, and the laptop on the right is the information carrier. Assuming the information is stored in the computer from time t1 to t2, any time in this interval can be considered as the reflection time duration of the carrier and the picture file, and a sequence of binary code stored in the laptop computer is the reflection set of the information.

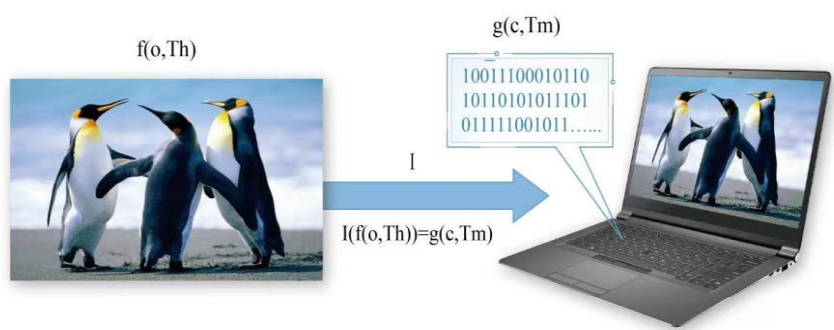

Fig. 1: Penguin activity information model on a laptop computer

Through this simple example, we can see that the information on the penguins' motions or perhaps even subjective emotions is stored in the computer as a static object in the objective world. The content of the information is the binary image file stored in the computer's memory, obeying all laws of science and technology in the objective physical world. This example illustrates the universal attribute of all information carried in modern information systems: it is independent of people's subjective will. Therefore, we call this philosophical understanding of information objective information theory (OIT).

On the other hand, by processing the binary image format file appropriately, the laptop computer can restore the image of three penguins under blue sky and white clouds on its display screen. It can be understood as the process of inverse mapping of the reflection set in a mathematical sense, recovering the state set of the information noumena. In this way, the most important attribute of information, namely, the ability to restore it to its original state, can be realized. If the storage size of this image file in the computer is 1 MB, then the commonly accepted information volume is simply 1 MB, with no need to estimate the probabilities of occurrence or the entropy formula. The 1 MB information volume will remain unchanged even after people become familiar with the picture.

People usually consider only the binary image file in a laptop as information, which means that only the reflection set in the carrier is recognized as information. This will not, in general, lead to serious problems since people have unconsciously formed in their mind the noumena that the "information" in the carriers represents. However, in a strict scientific sense, treating the set of reflections in the carrier as information itself will greatly limit our full study of information. As far as the example is concerned, if the binary image format file in the laptop is used as an independent object, we cannot study its authenticity, delay, and accuracy, which are all consequential key issues in the study of information. Therefore, only by incorporating the noumena and its carrier, together with their existence duration and states, into a complete model of information, can all characteristics of information be understood and a sound theoretical foundation be established to address many profound problems in the development of information science and technology.

**2.2 Sextuple model of information**

Based on the example just discussed and a basic understanding that information is parallel to matter and energy in its existence, to achieve a possible consensus on the nature of information and establish formal information models, we first propose four postulates about the concept of information:

**Postulate 1** (Binary Attribute) Any information consists of two parts: the noumena, which is the fundamental essence of the information, and the carrier, which bears the objective form of the information.

An important difference between information and matter or energy is that matter and energy can exist independently, but there is always a "shadow" lurking behind information. Questions are often raised concerning a specific piece of information: "Is this information true or false?", "How accurate is this information?". Indeed, the "surface" of the information is always compared with the "shadow" behind it, and the fundamental significance of information application is related to the ability to recover and apply the truth hidden behind the apparent outside form of information. We stress that "objective form" requires that the apparent outside form is not changed by people's subjective will. This requires that the carrier must be something that exists in the objective world. In the simple example of Section 1, people appreciate the state of three penguins' lives through the objective carrier, the laptop computer. This is the basis for Postulate 1.

**Postulate 2** (Existence Duration) The content and form of information exist in their own respective time periods.

Just as for matter and energy, the existence of a particular piece of information is obviously temporal, regardless of whether it is information's noumena or its objective form in the carrier. Information can be generated at a certain time or disappear at a certain time. The existence duration of the noumena and carrier has profound significance for information. For example, the study of information delay depends crucially on the relationship between the existence durations of the noumena and carrier. In the example in Section 1, for any time $t \in [t_1, t_2]$, $t - (t_0 + 0.01)$ is the delay of the image information. In many areas of information application, such as disastrous weather forecasting, military intelligence, and document exchange, this delay has an almost decisive influence. This is the basis for Postulate 2.

**Postulate 3** (State Representation) States of the noumena and the carrier of information have their own respective representation.

Another important difference between information and matter or energy is that the essence of the content of matter is still matter, and the essence of the content of energy is still energy. By contrast, the essence of the content of information is the state of events or objects. Furthermore, due to the binary attribute nature of information, neither state should be neglected. In the example of Section 1, the posture of the penguins is the state of the noumena of information; the binary image file in the laptop computer is the state of the carrier of information. We must compare the states of the noumena and the carrier if we wish to analyze the authenticity and accuracy of information. This is the basis for Postulate 3.

**Postulate 4** (Enabling Mapping) The state of the noumena can be mapped to the state of the carrier via enabling mapping. The implication of the existence of enabling mapping is that we can mathematically establish a surjective mapping from the state of the noumena to the state of the carrier and that the state of the carrier becomes objective reality due to the state of the noumena.

If there are no relations between the states of the noumena and the carrier, there is no information. Only when a surjective mapping relationship between two states is established mathematically and a cause-and-effect relationship can be formed from the former to the latter physically, does information come into existence. In the example in Section 1, the image information of penguins is precisely the mapping relationship between the Penguins' noumena state and the image file in the laptop computer established by information creators. The objective reality of the image file in the carrier comes into existence because of the existence of the state of the penguins' noumena. This is the basis for Postulate 4.

Postulates 1-4 axiomatize the necessary elements of information and relations among elements, from which the following sextuple model of information can be derived.

**Lemma 1** The necessary and sufficient condition for the existence of an information model satisfying Postulates 1-4 is the existence of an enabling mapping $I$ such that the set of states of noumena $f = f(o, T_h)$ is surjectively mapped to the set of carrier reflection states $g = g(c, T_m)$, where $o$ is a nonempty set of information noumena, $T_h$ is the set of times when noumena in $o$ occur, $c$ is a set of objective carriers, and $T_m$ is the set of reflection times of carriers.

We call map $I$, denoted by $I = \langle o, T_h, f, c, T_m, g \rangle$, the sextuple model of information.

**Proof** If Postulates 1-4 are satisfied, we have nonempty sets $o$ and $c$ by Postulate 1. Postulate 2 yields the sets $T_h$ and $T_m$, the time durations for the occurrences of noumena and reflections in carriers. Postulate 3 provides the set of states of noumena, denoted by $f(o, T_h)$, which depends on the set of noumena $o$ and the set of noumena occurrence times $T_h$, as well as the set of states of reflections in the carrier, denoted by $g(c, T_m)$, which further depends on $c$, the set of carriers, and $T_m$, the set of reflection times. According to Postulate 4, there exist surjective enabling mappings from $f(o, T_h)$ to $g(c, T_m)$. We denote such mappings by $I$ and obtain the sextuple model $I = \langle o, T_h, f, c, T_m, g \rangle$. The necessity is proven.

On the other hand, if the sets $o$, $c$, $T_h$, $T_m$, $f(o, T_h)$, and $g(c, T_m)$ exist together with a surjective enabling mapping from $f(o, T_h)$ to $g(c, T_m)$, then the sextuple model $I = \langle o, T_h, f, c, T_m, g \rangle$ clearly satisfies all four postulates. The sufficiency is also proven.

## 2.3 Definition of information and restorable information

Based on Postulates 1-4 and Lemma 1, we can now provide a mathematical definition of information in OIT.

**Definition 1** (Mathematical definition of information) Let $O$ denote the set of all content of the objective world, let $S$ denote the set of the content of the subjective world, and let $T$ be the set of times. The elements in these sets can be specified according to the requirements in different areas. A noumena is thus an element of the power set $2^{O \cup S}$ (or a subset of $O \cup S$), the occurrence duration $T_h \in 2^T$, and $f(o, T_h)$ is the set of states of $o$ over $T_h$. Carrier $c \in 2^O$, the set of reflection times $T_m \in 2^T$, and the set of reflection states $g(c, T_m)$ are all nonempty sets. Information $I$ is thus an enabling mapping from $f(o, T_h)$ to $g(c, T_m)$, i.e., $I: f(o, T_h) \to g(c, T_m)$ or $I(f(o,T_h)) = g(c,T_m)$. The collection of all information is called the information space, denoted by $\mathfrak{I}$, one of the three essential elements of the objective world.

Notably, the mapping $I$ must be a surjective map from $f(o,T_h)$ to $g(c,T_m)$ in a strict mathematical sense. However, $I$ can be called information only if $g(c,T_m)$ is created through $f(o,T_h)$ in the objective world either by self-excitation or external force. This is why the concept of enabling mapping is introduced in Definition 1. Therefore, a surjective map from $f(o,T_h)$ to $g(c,T_m)$ in a purely mathematical sense is not information if the occurrence of $g(c,T_m)$ has no physical connection to $f(o,T_h)$. Being surjective is a necessary, but not sufficient, condition for information. Information is indeed both mathematical and physical [30]; this is the essential property of information.

**Definition 2** (Restorable information) If information $I = \langle o, T_h, f, c, T_m, g \rangle$ is also an injective map from $f(o,T_h)$ to $g(c,T_m)$, i.e., for any $o_\lambda, o_\mu \in o, T_{h\lambda}, T_{h\mu} \in T_h, f_\lambda, f_\mu \in f$, when $f_\lambda(o_\lambda, T_{h\lambda}) \neq f_\mu(o_\mu, T_{h\mu})$, one must have $I(f_\lambda(o_\lambda, T_{h\lambda})) \neq I(f_\mu(o_\mu, T_{h\mu}))$. In this case, $I$ is an invertible mapping: its inverse $I^{-1}$ exists. For any $c_\lambda \in c, T_{m\lambda} \in T_m, g_\lambda \in g$, there exists a unique $o_\lambda \in o, T_{h\lambda} \in T_h, f_\lambda \in f$, such that $I^{-1}(g_\lambda(c_\lambda, T_{m\lambda})) = f_\lambda(o_\lambda, T_{h\lambda})$. Thus, $I^{-1}(g(c,T_m)) = f(o,T_h)$. We call information $I$ restorable, and $f(o,T_h)$ is the restored state of information $I$.

We see that from $g(c,T_m)$ and $I^{-1}$, we can restore the state of $o$ over $T_h$, $f(o,T_h)$. This is the restorability of information. In the real world, people search for the restored state through information. That is precisely the most important property and meaning of information.

## 3 Some corollaries of information metrics

The original intention of OIT was to enrich and improve measurement systems of information and support the efficacy analysis and research of information systems. Following the sextuple model of information, eleven types of information metrics can be concretely defined[26]. For each metric, a corresponding example can be found from classical or commonly used information principles.

### 3.1 The volume of information and Shannon's information entropy

The volume of restorable information $I = \langle o, T_h, f, c, T_m, g \rangle$ can be defined as a measure of its reflection set, i.e., $\text{volume}_\sigma(I) = \sigma(g(c, T_m))$.

**Corollary 1** (Minimum restorable volume of random event information) Let X be an event taking possible values $x_i$ randomly with the probability of $X = x_i$ equal to $p_i$, $i = 1, \ldots, n$. Assume that events $\{X = x_i\}$ and $\{X = x_j\}$ are independent when $i \neq j$ and $\sum_{i=1}^n p_i = 1$, and information $I = \langle o, T_h, f, c, T_m, g \rangle$ represents the code of X's values transmitted over a communication channel. In this case, $I$ is called random event information. In this information $I$, the noumena $o$ is the random event X, the occurrence duration $T_h$ is the occurrence duration of the event X, and the state set $f(o,T_h)$ consists of values $x_i$ and $i = 1, \ldots, n$. Carrier $c$ is the channel for transmitting the value of X, reflection duration $T_m$ is the time for channel $c$ to transmit the value of X, and reflection set $g(c,T_m)$ is the set of specific codes for channel $c$ to transmit the value X. If measure $\sigma$ represents the number of bits of $g(c,T_m)$, then the minimum volume for which $I$ can be restored is $\text{volume}_\sigma(I) = -\sum_{i=1}^n p_i \log_2 p_i$.

**Proof** As in [1], we assume that the semantics of communication are independent of engineering problems. Therefore, to minimize the required channel bandwidth, the communication process for transmitting the value of X does not need to directly transmit the specific value of $x_i$ but only needs to use different binary codes to represent the event that X takes the value of $x_i$, $i = 1, \ldots, n$ and then transmit the corresponding codes during the communication process. That is, as long as an appropriate coding method for $g(c, T_m)$ is selected, the required channel bandwidth can be reduced.

We now prove the corollary using proof by contradiction. Assume, on the contrary, that there is also some encoding of $g(c, T_m)$ such that $H' = \sigma(g(c, T_m)) < -\sum_{i=1}^n p_i \log_2 p_i = H$, and $I$ remains information that can be restored (i.e., the communication channel can transmit the source information completely), where $H$ is the information entropy of event X. Assume that the bandwidth of channel $c$ is $W$; then, we have $W/H' > W/H$, that is, channel $c$ can completely transmit the source information at a rate greater than $W/H$. This obviously contradicts the conclusion of Theorem 9 of [1] that the channel transmission rate $W/H$ cannot be exceeded.

Therefore, there is no encoding method such that $\sigma(g(c, T_m)) < -\sum_{i=1}^n p_i \log_2 p_i$, while keeping $I$ as restorable information. Additionally, $\text{volume}_\sigma(I) = -\sum_{i=1}^n p_i \log_2 p_i$ is the minimum information volume of $I$. The corollary is proved.

Corollary 1 shows that the information volume defined by OIT can also express Shannon's information entropy principle, and the information volume defined on the basis of information entropy is simply a special case of information volume. The definition of information volume requires only that $I$ is a one-to-one surjective mapping from $f(o, T_h)$ to $g(c, T_m)$ and that $g(c, T_m)$ is a measurable set with respect to measure σ. Compared with information entropy, this definition has much fewer mathematical constraints, so it should have broader applications.

## 3.2 The delay of the information and serial information transmission chain

The delay of restorable information $I = \langle o, T_h, f, c, T_m, g \rangle$ can be defined as the difference between the supremum of its reflection time and the supremum of its occurrence time, i.e., $\text{delay}(I) = \sup T_m - \sup T_h$.

According to the transitivity of information, we can let a set $\{I_i = \langle o_i, T_{hi}, f_i, c_i, T_{mi}, g_i \rangle | i = 1,\,,n\}$ be a set of restorable information[26]. If, for any $i < n$, we obtain $c_i = o_{i+1}, T_{mi} = T_{h(i+1)}, g_i = f_{i+1}$, then $\{I_i = \langle o_i, T_{hi}, f_i, c_i, T_{mi}, g_i \rangle | i = 1,\,,n\}$ is the serial information transmission chain between information $I_1$ and $I_n$, and there exists $\{I_i^{'} = \langle o_1, T_{h1}, f_1, c_i, T_{mi}, g_i \rangle | i = 1,\,,n\}$, which are both restorable information and have the same restored state $f_1(o_1, T_{h1})$.

**Corollary 2** (Serial information transfer delay) Let $\{I_i = \langle o_i, T_{hi}, f_i, c_i, T_{mi}, g_i \rangle | i = 1,\,,n\}$ be the serial information transmission chain between information $I_1$ and $I_n$. Then, $I = \langle o_1, T_{h1}, f_1, o_n, T_{mn}, g_n \rangle$ is also restorable information, and $\text{delay}(I)$ is the sum of the delays of all information $I_i$.

**Proof** For the serial information transmission chain $\{I_i = \langle o_i, T_{hi}, f_i, c_i, T_{mi}, g_i \rangle | i = 1,\,,n\}$, by its definition, for any $i < n$, we have $T_{mi} = T_{h(i+1)}$. Therefore, we have $\sum_{i=1}^n \text{delay}(I_i) = \sum_{i=1}^n (\sup T_{mi} - \sup T_{hi}) = \sup T_{mn} - \sup T_{hn} + \sum_{i=1}^{n-1}(\sup T_{h(i+1)} - \sup T_{hi}) = \sup T_{mn} - \sup T_{h1} = \text{delay}(I)$. The corollary is proved.

## 3.3 The scope of information and the radar equation

The scope of restorable information $I = \langle o, T_h, f, c, T_m, g \rangle$ can be defined as the measure of its noumena, i.e., $\text{scope}_\sigma(I) = \sigma(o)$.

**Corollary 3** (Extent of radar detection information) Let restorable information $I =$

$\langle o, T_h, f, c, T_m, g \rangle$ be the radar detection information, the noumena $o$ be the object detected by the radar, the state occurrence time $T_h$ be the time when the radar beam shines on the detected object, and state set $f(o, T_h)$ be the state of the detected object itself and its motion. In this case, carrier $c$ is the radar, the reflection time $T_m$ is the time required for the radar to receive, process, store, and display the echo signal and/or the data of the detected object, and reflection set $g(c, T_m)$ is the echo signal and/or data of the detected object for the radar to receive, process, store, and display. In addition, we define measure $\sigma$ of noumena $o$ as its reflection area. According to the radar equation, when the radar transmitting power, antenna gain, antenna effective aperture, and minimum detectable signal are determined, the maximum detection range of the radar depends on the $\text{scope}_\sigma(I)$ of information $I$ and is proportional to its quartic root.

**Proof** According to the definition of radar detection information $I = \langle o, T_h, f, c, T_m, g \rangle$, noumenon $o$ is the object detected by radar, and the $\text{scope}_\sigma(I)$ of $I$ is the reflection area $\sigma$ of $o$. In the radar equation $R_{max}^4 = P_t G_t A_e \sigma / (4\pi)^2 S_{min}$[3], $R_{max}$ is the maximum detection range of the radar, $P_t$ is the radar transmitting power, $G_t$ is the radar antenna gain, $A_e$ is the effective aperture of the radar antenna, $S_{min}$ is the minimum detectable signal of the radar, and $\sigma$ is the reflection area of the detected object. When the important parameters $P_t$, $G_t$, $A_e$, and $S_{min}$ of the radar itself are determined, the maximum detection range of the radar is determined entirely by $\sigma$ and is proportional to the fourth root of the scope of information $I$. The corollary is proved.

### 3.4 The granularity of information and Rayleigh criterion for optical imaging

The granularity of restorable information $I = \langle o, T_h, f, c, T_m, g \rangle$ can be defined as the average of all noumenon measures of its atomic information, i.e., $\text{granularity}_\sigma(I) = \int_\Lambda \sigma(o_\lambda) d\mu / \mu(\Lambda)$, where $o_\lambda (\lambda \in \Lambda)$ is a noumenon of the atomic information, $\Lambda$ is the index set, and $\mu$ is the measure of $\Lambda$.

**Corollary 4** (Resolution of optical imaging information) Let the restorable information $I = \langle o, T_h, f, c, T_m, g \rangle$ be the optical imaging information, the noumena $o$ be the object to be photographed or filmed, the state occurrence time $T_h$ be the time when the shutter is opened or the camera acquires the videos, and the state set $f(o, T_h)$ be the state of the photographed object itself and its motion. In this case, carrier $c$ is an image or video in the camera, reflection time $T_m$ is the time required by the camera to shoot, process, store, and display the image of the target object, and reflection set $g(c, T_m)$ is the image or video of the target object. Here, we define the measure $\sigma$ of noumena $o$ as the minimum distinguishable angle at the time the target object is photographed; then, the resolution of the optical imaging information $I$, i.e., $\text{granularity}_\sigma(I)$, is proportional to the wavelength of light and inversely proportional to the width of the photosensitive unit.

**Proof** For an object $o$ to be shot, each frame in the optical imaging information $I = \langle o, T_h, f, c, T_m, g \rangle$ contains a large number of pixel points, and each pixel point is a local image of $o$ that cannot be subdivided. Thus, the pixel represents the atomic information $I_\lambda = \langle o_\lambda, T_{h\lambda}, f_\lambda, c_\lambda, T_{m\lambda}, g_\lambda \rangle$, where $\lambda \in \Lambda$, and $\Lambda$ is the index set. By definition, the $\text{granularity}_\sigma(I)$ of information $I$ is the average of the measure $\sigma(o_\lambda)$ of the noumena $o_\lambda$ of all atomic information $I_\lambda$. The principle of optical imaging, indicates that $\sigma(o_\lambda)$ is the same for all $\lambda \in \Lambda$, and according to the Rayleigh criterion[31], $\sigma(o_\lambda) = l/a$, where $l$ is the wavelength of light and $a$ is the width of the photosensitive unit. Thus, the $\text{granularity}_\sigma(I) = \sigma(o_\lambda)$ (for any $\lambda \in \Lambda$) of information $I$ is proportional to the wavelength of light and inversely proportional to the width of the photosensitive unit. The corollary is proved.

## 3.5 Invariance principle of restorable information variety

The variety of restorable information $I = \langle o, T_h, f, c, T_m, g \rangle$ can be defined as the number of equivalence classes on its set of states, i.e., $\text{vriety}_R(I) = \overline{[f(o, T_h)]_R}$, where $R$ is an equivalence relation on the set of states $f(o, T_h)$.

**Corollary 5** (Invariance of restorable information variety) For restorable information $I = \langle o, T_h, f, c, T_m, g \rangle$, let $R$ be the equivalence relation on the state set $f(o, T_h)$; the set of equivalence classes of the elements in $f(o, T_h)$ relative to $R$ is $[f(o, T_h)]_R$. Then, there must exist an equivalence relation $Q$ on the reflection set $g(c, T_m)$ such that set $[g(c, T_m)]_Q$, the set of equivalence classes of the elements in $g(c, T_m)$ relative to $Q$, and set $[f(o, T_h)]_R$ form a one-to-one surjective relation by information $I$ so that the cardinalities of the two equivalence classes are equal. This is expressed as $\text{vriety}_R(I) = \overline{[f(o, T_h)]_R} = \overline{[g(c, T_m)]_Q}$.

**Proof** It is easy to prove that the equivalence relation $Q$ on reflection set $g(c, T_m)$ can be established according to the equivalence relation $R$ on the state set $f(o, T_h)$. For any two pieces of subinformation of $I$, $I_\lambda = \langle o_\lambda, T_{h\lambda}, f_\lambda, c_\lambda, T_{m\lambda}, g_\lambda \rangle$ and $I_\mu = \langle o_\mu, T_{h\mu}, f_\mu, c_\mu, T_{m\mu}, g_\mu \rangle$ if $f_\lambda(o_\lambda, T_{h\lambda}) R f_\mu(o_\mu, T_{h\mu})$, there must be $g_\lambda(c_\lambda, T_{m\lambda}) Q g_\mu(c_\mu, T_{m\mu})$. Thus, there is also an equivalence class $[g(c, T_m)]_Q$ on $g(c, T_m)$. Note that the equivalence relation $Q$ is entirely established based on the mapping relation of information $I$. Thus, the one-to-one surjective relation between the two equivalence classes $[f(o, T_h)]_R$ and $[g(c, T_m)]_Q$ can also be established entirely based on information $I$. Hence, we have $\overline{[f(o, T_h)]_R} = \overline{[g(c, T_m)]_Q}$. The corollary is proved.

## 3.6 Average duration of continuous monitoring information

The duration of restorable information $I = \langle o, T_h, f, c, T_m, g \rangle$ can be defined as the difference between the supremum and infimum of $T_h$, i.e., $\text{duration}(I) = \sup T_h - \inf T_h$.

**Corollary 6** (Average duration of continuous monitoring information) The average duration of continuous monitoring information is equal to the MTBF of the information collection device.

**Proof** Let restorable information $I = \langle o, T_h, f, c, T_m, g \rangle$ be the continuous monitoring information, where the noumena $o$ can be considered the monitored object, the state occurrence time $T_h$ can be considered the time period when $o$ is in the monitored state, the state set $f(o, T_h)$ can be considered the state when $o$ is in the monitored time period, and carrier $c$ can be considered the information collection device. In addition, the reflection time $T_m$ can be considered the working period of $c$, and the reflection set $g(c, T_m)$ can be considered the information collected and presented by $c$. A continuous monitoring information system is generally required to maintain continuous and uninterrupted monitoring of the target object, i.e., its duration is often equal to the working period of the information collection equipment. Thus, we have $\text{duration}(I) = \sup T_h - \inf T_h = \sup T_m - \inf T_m$.

However, any equipment has a possibility of failure; therefore, it is necessary to specify the mean fault-free MTBF [32] of systems in engineering practice, which indicates the duration of the normal operation of the system without failure over the entire life cycle. In continuous monitoring systems, since the MTBF of information acquisition equipment $c$ is the average of all the $\sup T_m - \inf T_m$ values in the full life cycle, the average of the continuous monitoring information $\text{duration}(I)$ is the same. The corollary is proved.

## 3.7 Sampling_rate of information and Nyquist's sampling theorem

The sampling_rate of restorable information $I = \langle o, T_h, f, c, T_m, g \rangle$ can be defined as the ratio of the number of interruptions of its occurrence time to the length of the same occurrence time, i.e., sampling_rate$(I) = \bar{\Lambda}/|U|$, where $U = \bigcup_{\lambda \in \Lambda} U_\lambda$, $U_\lambda \subseteq [\inf T_h, \sup T_h]$ and $T_h \cap U_\lambda = \emptyset$, and $\Lambda$ is an index set.

**Corollary 7** (Minimum restorable sampling_rate of periodic information) For restorable information $I = \langle o, T_h, f, c, T_m, g \rangle$, if $f(o, T_h)$ is a numeric set and there is a minimum of $T$ such that, for $\forall x \in o, t, t+T \in T_h$, $f(x,t) = f(x, t+T)$, then $I$ is referred to as the periodic information, and the lowest restorable sampling_rate of information $I$ is equal to $1/(2T)$.

**Proof** For periodic information $I = \langle o, T_h, f, c, T_m, g \rangle$, $\inf T_h \neq \sup T_h$. Here, $f(x,t) = f(x, t+T)$ for $\forall x \in o, t, t+T \in T_h$, and $T$ is the smallest value satisfying this condition. Thus, for $\forall x \in o$, $f(x,t)$ contains no frequency greater than $1/T$ relative to time $t$. According to Nyquist's sampling theorem [5], $f(x,t)$ is completely determined relative to time $t$ by a series of values that are no greater than $T/2$ apart. In the definition of information $I$, $\{U_\lambda\}_{\lambda \in \Lambda}$ is a series of sampling intervals with equal measure, and the cardinal number $\bar{\Lambda}$ is the number of sampling intervals. Therefore, the Lebesgue measure $|U| = |U_\lambda|\bar{\Lambda}$ of $U = \bigcup_{\lambda \in \Lambda} U_\lambda$ holds for $\forall \lambda \in \Lambda$, and the sampling_rate of information $I$ sampling_rate$(I) = \bar{\Lambda}/|U| = \bar{\Lambda}/|U_\lambda|\bar{\Lambda} = 1/|U_\lambda|$ holds for $\forall \lambda \in \Lambda$.

Note that $|U_\lambda| \leq T/2$ holds for $\forall \lambda \in \Lambda$ if and only if sampling_rate$(I) \geq 1/(2T)$. Thus, for $\forall x \in o$, the values of $f(x,t)$ are determined completely, and $I$ has a definite restored state. The corollary is proved.

### 3.8 Invariance principle of the aggregation degree of restorable information

The aggregation of restorable information $I = \langle o, T_h, f, c, T_m, g \rangle$ can be defined as the ratio of the number of relations between all elements in the state set $f(o, T_h)$ to the number in the set $f(o, T_h)$, i.e., aggregation$(I) = \bar{\Re}/\overline{f(o, T_h)}$, where $\Re$ is the set of relations between all elements in the state set $f(o, T_h)$.

**Corollary 8** (Invariance of the aggregation degree of restorable information) For restorable information $I = \langle o, T_h, f, c, T_m, g \rangle$, let the cardinality of the state set $f(o, T_h)$ be $\overline{f(o, T_h)} \neq 0$, and let $\Re$ be the set of all relations between the elements of $f(o, T_h)$. Then, for the reflection set of $g(c, T_m)$, its cardinality $\overline{g(c, T_m)} = \overline{f(o, T_h)}$, and there exists a set of relations $\mathfrak{Q}$ on it such that cardinality $\bar{\mathfrak{Q}} = \bar{\Re}$. Thus, we have aggregation$(I) = \bar{\Re}/\overline{f(o, T_h)} = \bar{\mathfrak{Q}}/\overline{g(c, T_m)}$.

**Proof** Here, $I = \langle o, T_h, f, c, T_m, g \rangle$ is restorable information; thus, there is a surjective map from the state set $f(o, T_h)$ to the reflection set $g(c, T_m)$. Therefore, the cardinalities of the two sets are necessarily equal, i.e., $\overline{g(c, T_m)} = \overline{f(o, T_h)}$. In addition, we can define the set of relations $\mathfrak{Q} = \{Q_R | R \in \Re\}$ on $g(c, T_m)$ such that, for any $R \in \Re$, if $o_\lambda, o_\mu \in o$, $T_{h\lambda}, T_{h\mu} \in T_h$, $f_\lambda, f_\mu \in f$, and $f_\lambda(o_\lambda, T_{h\lambda}) R f_\mu(o_\mu, T_{h\mu})$, then $I(f_\lambda(o_\lambda, T_{h\lambda})), I(f_\mu(o_\mu, T_{h\mu})) \in g(c, T_m)$, and we define $f_\lambda(o_\lambda, T_{h\lambda}) Q_R f_\mu(o_\mu, T_{h\mu})$ such that $Q_R$ is the relation on $g(c, T_m)$. The set $\mathfrak{Q} = \{Q_R | R \in \Re\}$ has a one-to-one surjective relationship with set $\Re$; thus, the cardinalities are exactly equal, i.e., $\bar{\mathfrak{Q}} = \bar{\Re}$. From this result, we obtain aggregation$(I) = \bar{\Re}/\overline{f(o, T_h)} = \bar{\mathfrak{Q}}/\overline{g(c, T_m)}$. The corollary is proved.

### 3.9 Scope and coverage of information and Metcalfe's law

The coverage of restorable information $I = \langle o, T_h, f, c, T_m, g \rangle$ can be defined as the integral of all measures of the carriers of its copies and itself, i.e., coverage$_\sigma(I) = \int_\Lambda \sigma(c_\lambda) d\mu$, where $\{I_\lambda = \langle o_\lambda, T_{h\lambda}, f_\lambda, c_\lambda, T_{m\lambda}, g_\lambda \rangle\}_{\lambda \in \Lambda}$ is a set containing $I$ and all its copies, $\Lambda$ is an index set, $\mu$ is a measure on index set $\Lambda$, and $\sigma$ is a measure on measurable set $c$.

**Corollary 9** (The value of a network system is equal to the product of the maximum scope and the maximum coverage of the information it carries) For restorable information $I = \langle o, T_h, f, c, T_m, g \rangle$, if its carrier $c$ is a network system comprising finite nodes, its value is equal to the product of the maximum possible values of the scope and coverage of $I$.

**Proof** Given that carrier $c$ in restorable information $I = \langle o, T_h, f, c, T_m, g \rangle$ is a network system comprising a finite number of nodes, if the number of nodes in $c$ is $n$, according to Metcalfe's law [7], the value of the network system $c$ is equal to the square of the number of nodes $n^2$. In addition, we can consider $I$ as the information from all nodes in a network; thus, the noumena $o$ is the network system $c$, and its measure $\sigma$ is the number of nodes. In this case, the maximum scope of information $I$ is $\text{scope}_\sigma(I) = n$. The measure of carrier $c$ is also the number of nodes, and the maximum value of the coverage of information $I$ is $\text{coverage}_\sigma(I) = n$. Therefore, the value of this network system is equal to the product of the maximum possible values of the scope and coverage of information $I$. The corollary is proved.

### 3.10 Distortion of information and Kalman filtering principle

The distortion of restorable information $I = \langle o, T_h, f, c, T_m, g \rangle$ can be defined as the distance between its reflected state and its restored state in distance space, i.e., $\text{distortion}_J(I) = d(f, \tilde{f})$, where $\tilde{f}(\tilde{o}, \widetilde{T_h})$ is its reflected state.

**Corollary 10** (Minimum distortion estimation method for discrete linear stochastic systems) Let $I = \langle o, T_h, f, c, T_m, g \rangle$ be the state information of a discrete linear stochastic system, of which the motion and measurement are both affected by Gaussian white noise. Then, the minimum distortion estimation of $I$ can be obtained based on the reflection $J$ of the Kalman filter [9].

**Proof** For the state information $I = \langle o, T_h, f, c, T_m, g \rangle$ of a discrete linear stochastic system in which the motion and measurement are both affected by Gaussian white noise, the noumena $o$ is the system itself, the set of occurrence times $T_h$ is a series of time sequences with equal intervals, which can be written as $1,2,,,k$, the state set $f(o, T_h)$ can be written as $x(k), k = 1,2,\cdots$, and $x(k) = Ax(k-1) + BU(k) + W(k)$ holds. Here, $x(k)$ is the system state at time $k$, and $U(k)$ is the system input at time $k$. $A$ and $B$ are system parameters, which are matrices for a multimodel system, $W(k)$ is the motion noise of the system, and $Q$ is its covariance.

In this case, carrier $c$ is a measuring system, and the reflection time set $T_m$ is the same as the occurrence time set $T_h$ (denoted $1,2,\cdots k,\cdots$). The reflection set $g(c, T_m)$ is a series of measured values on $T_m$, which is denoted by $z(k), k = 1,2,....,$ and $z(k) = Hx(k) + V(k)$ holds, where $z(k)$ is the measured value at time $k$. In addition, $H$ is the measuring system parameter, and for a multimeasurement system, $H$ is a matrix, $V(k)$ is the measuring noise at time $k$, and $R$ is its covariance. If $J$ consists of the following five formulas

$$x(k|k-1) = Ax(k-1|k-1) + BU(k),$$

where $x(k|k-1)$ is the result predicted by the previous state, $x(k-1|k-1)$ is the optimal result of the previous state, and $U(k)$ is the current system input state;

$$P(k|k-1) = AP(k-1|k-1)A^T + Q,$$

where $P(k|k-1)$ is the covariance corresponding to $x(k|k-1)$, $P(k-1|k-1)$ is the covariance corresponding to $x(k-1|k-1)$, $A^T$ is the transpose of matrix $A$, and $Q$ is the covariance of the motion of the system;

$$x(k|k) = x(k|k-1) + G(k)(z(k) - Hx(k|k-1))$$

where $G(k)$ is the Kalman gain;

$$G(k) = P(k|k-1)H^T/(HP(k|k-1)H^T + R)$$
$$P(k|k) = (I - G(k)H)P(k|k-1)$$

where $I$ is the identity matrix;

then, it is obvious that $J$ can be inferred recursively to be the surjective map from $z(k)$ to $x(k)$, i.e., from $g(c, T_m)$ to $f(o, T_h)$. Thus, $J$ is a reflection of $I$. Consequently, according to the Kalman filtering principle [9], $x(k|k)$ is the optimal estimation of $x(k)$, i.e., the minimum distortion estimation of $I$ can be obtained based on reflection $J$ of the Kalman filter. The corollary is proved.

### 3.11 Mismatch of information and average lookup length of a search algorithm

The mismatch of restorable information $I = \langle o, T_h, f, c, T_m, g \rangle$ can be defined as the distance between the target information and itself in distance space, i.e., $mismatch_{I_0}(I) = d(I, I_0)$, where $I_0$ is the target information.

**Corollary 11** (Average search length of minimum mismatch information for search algorithms) Let the target information $I_0 = \langle o_0, T_{h0}, f_0, c_0, T_{m0}, g_0 \rangle$ and set $\{I_i = \langle o_i, T_{hi}, f_i, c_i, T_{mi}, g_i \rangle | i = 1,,,n\}$ be restorable information. Here, $o_0$ and $o_i$, $T_{h0}$ and $T_{hi}$, $f_0$ and $f_i$, $c_0$ and $c_i$, $T_{m0}$ and $T_{mi}$, and $g_0$ and $g_i$ are elements of sets $\mathcal{P}_o, \mathcal{P}_{T_h}, \mathcal{P}_f, \mathcal{P}_c, \mathcal{P}_{T_m}$, and $\mathcal{P}_g$, respectively, and $I_0$ and $I_i(i = 1,,,n)$ are elements of the distance space $\langle (\mathcal{P}_o, \mathcal{P}_{T_h}, \mathcal{P}_f, \mathcal{P}_c, \mathcal{P}_{T_m}, \mathcal{P}_g), d \rangle$. Let $1 \leq m \leq n$ such that the following holds.

$$mismatch_{I_0}(I_m) < mismatch_{I_0}(I_i), \quad 1 \leq i \leq n \text{ and } i \neq m$$

Then, the ASL from $\{I_i | i = 1,,,n\}$ to $I_m$ is related to both the mismatch $mismatch_{I_0}(I_m)$ and different search algorithms.

**Proof** With the increasingly abundant information content available on the internet and the increasing number of information query tools in use, a large number of complex information queries has arisen. Note that users have requirements for the noumena, occurrence time, state of information, carrier, reflection time, and mode of information. However, it is difficult to clearly describe all these requirements. Thus, target information that completely matches the users' requirements is difficult to find. Advanced retrieval or intelligent recommendation systems often analyze and estimate the target information $I_0$ that satisfies a user's needs for a given application scenario, and such systems search and calculate information $I_m$ with the minimum degree of mismatch $mismatch_{I_0}(I_i)(i = 1,,,n)$ from the limited information set $\{I_i | i = 1,,,n\}$, which is finally pushed to the end users.

According to the ASL principle [33], the definition of ASL is given as $ASL = \sum_{i=1}^{n} p_i c_i$, where $p_i$ is the probability of finding information $I_i$. Generally, we assume that the probability of finding each piece of information is the same, i.e., $p_i = 1/n$. Here, $c_i$ is the number of comparisons required to find the information $I_i$. Then, the following two situations must be considered.

The first situation is $mismatch_{I_0}(I_m) = 0$. Here, if the sequential search method is used, the mismatch degree $mismatch_{I_0}(I_i)$ is calculated incrementally from information $I_1$ until $I_m$ is found. Thus, we obtain $ASL = \sum_{i=1}^{n} p_i c_i = (1/n) \sum_{i=1}^{n} i = (n+1)/2$.

If the bisection search method is adopted, the middle serial number information is always used as the root to divide the left and right subtrees, and each subtree is used as the root of the middle serial number information to continuously divide in a progressive manner until the subtrees cannot be divided further. For each subtree, information $I_i$ is searched from the root, and the mismatch $mismatch_{I_0}(I_i)$ is calculated until $I_m$ is found. Thus, we obtain $ASL = \sum_{i=1}^{n} p_i c_i = (1/n) \sum_{i=1}^{n} 2^{i-1} i = ((n+1)/n) log_2(n+1) - 1$. Here, $h = log_2(n+1)$ is the height of the $n$ information discrimination trees.

The second situation is $mismatch_{I_0}(I_m) \neq 0$. In this case, the ASL is always $n$ because we must compare the mismatch degree $mismatch_{I_0}(I_i)(i = 1,,,n)$ of all information and select the minimum mismatch degree to obtain $I_m$. When $n$ takes a very large value, the amount of search computation is also very large due to ASL. Thus, an appropriate threshold can be set, and the search is completed when the mismatch degree $mismatch_{I_0}(I_i)$ is less than or equal to the threshold. The corollary is proved.

Therefore, we have demonstrated that for each metric defined by OIT, a corresponding example can be found in classical or common information theories and principles listed in Table 1, which also shows that OIT can accommodate many existing information scientific theories and technical methods.

Table 1: Relevant corollaries on metrics consistent with classic and commonly used principles[26]

| Metrics | Classic/common theories | Basic conclusion |
| --- | --- | --- |
| Volume | Shannon information entropy | The minimum restorable volume of random event information is its information entropy. |
| Delay | Serial information transmission delay | The overall delay of serial information transmission is equal to the sum of delays of all links. |
| Scope | Radar equation | The maximal detection distance is proportional to the fourth root of information scope. |
| Granularity | Rayleigh criterion for optical imaging | The granularity of optical imaging information is proportional to the wavelength of light and inversely proportional to the width of the sampling pore. |
| Variety | Invariance principle of restorable information variety | Restorable information can preserve the variety of information. |
| Duration | Average duration of continuous monitoring information | The average time duration of continuous monitoring information is equal to the mean time between consecutive failures of the information collection system. |
| sampling_rate | Nyquist sampling theorem | The lowest restorable sampling_rate of periodic information is equal to half the highest frequency of the state of the noumena. |
| Aggregation | Invariance principle of the aggregation degree of restorable information | Restorable information can preserve the aggregation degree of information. |
| Coverage | Metcalfe's law | The value of a network system is equal to the product of the maximum scope and the maximum coverage of all information that it carries. |
| Distortion | Kalman filtering principle | Kalman filtering is a minimum distortion estimation method for discrete linear random systems. |
| Mismatch | Average lookup length of a search algorithm | The average lookup length for the minimum mismatch information in a finite set of information is the average lookup length of a search algorithm. |

## 4 Relationship of information and matter, energy and time

Since matter, energy and information are the three elements of the objective world, and time is the necessary condition for carrying the three, the relationship between them naturally becomes a major basic issue of scientific research. OIT defines the volume of information as a measure of its reflection set [26], which, like mass, energy and time, is the most appropriate metric to express the relationship between these elements.

### 4.1 Combination of atomic information and volume additivity

The representation of information as a sextuple model $I = \langle o, T_h, f, c, T_m, g \rangle$ does not emphasize its nature as a set. We simply need to modify the notation slightly to see that, i.e., $I = \{\langle o_\lambda, T_{h\lambda}, f_\lambda, c_\lambda, T_{m\lambda}, g_\lambda \rangle\}$, where $o_\lambda \in o, T_{h\lambda} \in T_h, f_\lambda(o_\lambda, T_{h\lambda}) \in f(o, T_h), c_\lambda \in c, T_{m\lambda} \in T_m, g_\lambda(c_\lambda, T_{m\lambda}) \in g(c, T_m)$, $\lambda \in \Lambda$, an index set. We then see that $I$ is in fact a set of elements with six components. The concepts of subinformation and proper subinformation can be defined analogously based on those of subsets and proper subsets from set theory. Just as matter can be decomposed to indivisible elementary particles and energy can be decomposed to indivisible quanta, any information can also be decomposed to the most basic level at which further decomposition is no longer possible. This is the concept of atomic information (Figure 2). Atomic information does not mean that information takes an atom as its noumena or carrier, it is simply the most minute and basic component of the whole information space. Therefore, it plays a very important role in the study of the composition and nature of information. Any information can be regarded as a combination of all its atomic information, and there is no overlap among atomic information. Based on the additivity of a set and measure [34], we define the additivity of the volume of atomic information and its combination:

When $g_\lambda(c_\lambda, T_{m\lambda})$ is σ-measurable for any atomic information $I_\lambda = \langle o_\lambda, T_{h\lambda}, f_\lambda, c_\lambda, T_{m\lambda}, g_\lambda \rangle$ of restorable information $I = \langle o, T_h, f, c, T_m, g \rangle$, where $\lambda \in \Lambda$, a countable index set, then we have

$$\text{volume}_\sigma(I) = \sum_{\lambda=1}^{\infty} \text{volume}_\sigma(I_\lambda). \tag{1}$$

When $\Lambda$ is a finite set, we can simply change the superscript $\infty$ in the sum (1) to a finite number representing $\Lambda$'s cardinality. However, if $\Lambda$ is an uncountable set, formula (1) cannot be simply changed into an integral since, in general, the measure does not have additivity with regard to an uncountable set.

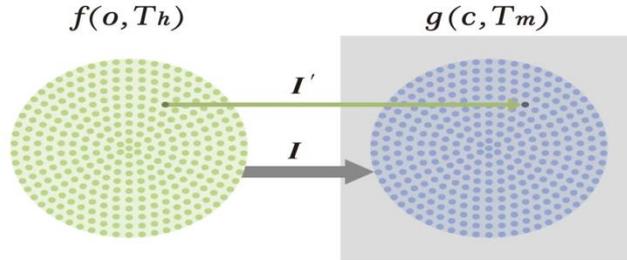

Fig. 2: Illustration of the concept of atomic information

### 4.2 Information volume that an individual quantum can carry

Quantum information theory is an important achievement of contemporary information technology[35][36]. It is commonly agreed that there is no radiation with energy smaller than a photon [37]. A single quantum such as a fermion (e.g, a quark or lepton) or boson (e.g., gluon, w-boson, photon, or graviton) [38], is a basic element that cannot be further divided in terms of energy; thus, it is the most efficient carrier of atomic information.

**Corollary 12** (Information volume of a single quantum carrier) For restorable information $I = \langle o, T_h, f, c, T_m, g \rangle$, when carrier $c$ is a single quantum, $I$ is called the single quantum carrier information. Define measure $\sigma$ as the number of distinguishable states experienced by $c$ over reflection time duration $T_m$; let $t = \sup T_m - \inf T_m$ be the length of the reflection time duration of information $I$. Then, the upper bound of the volume of information $I$ with respect to measure $\sigma$ is

$$\text{volume}_\sigma(I) = \begin{cases} 4\Delta E t/h, & t \text{ is sufficiently large;} \\ 1, & t \text{ is or is approaching } 0. \end{cases} \quad (2)$$

where the unit of $\text{volume}_\sigma(I)$ is a qubit, $\Delta E$ is the average energy of quantum $c$, and $h \approx 6.6 \times 10^{-34} \text{j/s}$ is the Planck constant.

**Proof** According to the definition of information volume, $\text{volume}_\sigma(I)$ is the number of distinguishable states experienced by quantum carrier $c$ over reflection time duration $T_m$. The state of a quantum is composed of two mutually orthogonal basis states $|0\rangle$ and $|1\rangle$ and their quantum superposition $a|0\rangle + b|1\rangle$, where $a$ and $b$ are complex numbers satisfying $|a|^2 + |b|^2 = 1$ [39]. However, distinguishable states of a quantum must be orthogonal [40]. Thus, by the Margolus-Levitin theorem [41], the shortest time delay from one state to another orthogonal state depends on the associated average energy $\Delta E$. The transition time is no shorter than $\Delta t = h/4\Delta E$. Since each distinguishable state of a single quantum can carry exactly one qubit of information [42][43][44], for carrier $c$, if the average energy is $\Delta E$, then for a length of reflection time duration $t$, if $t < \Delta t$, $c$ can display one only state that is distinguishable from others. This means $\text{volume}_\sigma(I) = 1$ (qubit), i.e., $c$ can carry only one piece of qubit information. This proves the second case of (2).

More generally, for any longer reflection time duration $t$, we have $\text{volume}_\sigma(I) = [t/\Delta t] + 1 = [4\Delta E t/h] + 1$, where $[\cdot]$ denotes the taking-the-integer-part function. Since $h \approx 6.6 \times 10^{-34}$, when $t$ is sufficiently large, for example, $\Delta E t \geq 10^{-30}$, by the proceeding formula, the relative error between $[4\Delta E t/h] + 1$ and $4\Delta E t/h$ is smaller than the scale of $10^{-4}$ and is negligible. Thus, we obtain $\text{volume}_\sigma(I) \approx 4\Delta E t/h$ (qubit).

Corollary 12 demonstrates that as a quantum carrier is limited by its associated energy, it can carry only a finite amount of different information over any time interval. In addition, as the number of quanta in the universe is finite and the information on a quantum carrier is the most basic component of information, even if time is assumed to be infinite in the future, mathematically, it can be guaranteed that any information is composed of at most countably much atomic information. Thus, under any circumstances, the additivity of volume for information combination will surely hold.

A single quantum is indivisible. Therefore, its state as a carrier at any moment is a reflection set of atomic information. The most straightforward case is information $I = \cup_{i=1}^n \{I_i = \langle c, T_{mi}, g, c, T_{mi}, g \rangle | i = 1, 2, \ldots n\}$, where $n$ is the number of orthogonal states experienced over time sequence $T_{mi}(i = 1, 2, \ldots n)$ by a single quantum $c$ and $g(c, T_{mi})$ is the state set of $c$ over time $T_{mi}$. Each $I_i$ represents atomic information and is a self-mapping from $f(o, T_{hi})$ to $g(c, T_{mi}) = f(o, T_{hi})$. Figure 3 illustrates that information $I$ reflects the transition situation of a single quantum $c$ state in the objective world, where $|0\rangle$ and $|1\rangle$ are two orthogonal basis states of quantum $c$. In this case, according to formula (1), $\text{volume}_\sigma(I) = \sum_{i=1}^n \text{volume}_\sigma(I_i) = n$ (qubit), and the upper bound of $n$ is given in formula (2).

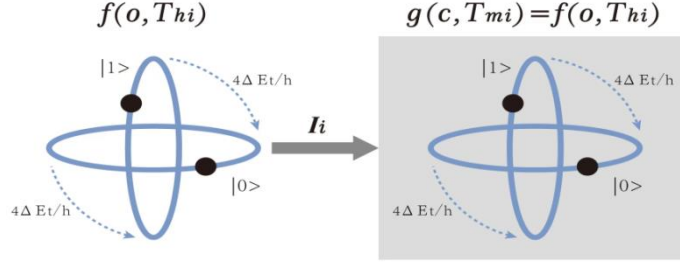

Fig. 3: Illustration of the self-mapping of the state of quantum information

More generally, information $I = \bigcup_{i=1}^n \{I_i = \langle o, T_{hi}, f, c, T_{mi}, g \rangle | i = 1,2, \ldots n\}$, where $n, c, T_{mi}, g(c, T_{mi}), i$ is as defined before, and $f(o, T_{hi})(i = 1,2, \ldots n)$ is the set of states experienced by the noumena $o$ in the subjective or objective world during the occurrence duration sequence $T_{hi}$. Clearly, each $I_i$ is also atomic information, and their combined information $I$ reflects the state of other noumena onto the state of a single quantum $c$ (Figure 4) so that we can use quantum information technology to efficiently process various items and phenomena in the subjective or objective world. In this case, we also have $\text{volume}_\sigma(I) = \sum_{i=1}^n \text{volume}_\sigma(I_i) = n$ (qubit), and the upper bound is given in formula (2).

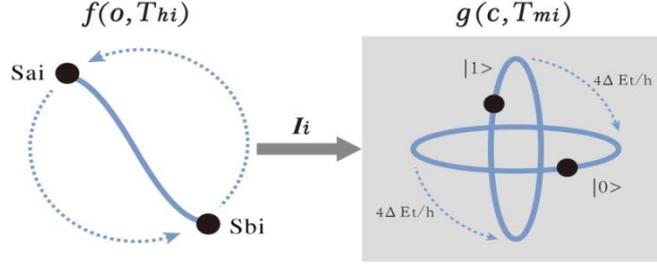

Fig. 4: Illustration of the transformation of general quantum information states

**4.3 Information volume that a general carrier can carry**

Since the discovery of cosmic expansion by Hubble [45], the standard assumption has been that all energy in the universe is in the form of radiation and ordinary matter[46]. Any matter or radiation in the objective world can be a carrier of information, and the information volume a carrier can carry depends on its physical properties and the level of technology [47]. At present, silicon chips are commonly the carriers with the largest carrying volume in information systems. With the most advanced manufacturing technology, the mass of a silicon chip is approximately 1.6 grams, and its storage volume can reach $10^{12}$ bits. Therefore, the information volume that a kilogram of silicon chips can carry is $6.25 \times 10^{14}$ bits.

According to principles of thermodynamics and the mass–energy conversion formula [48], the minimum mass of matter required to store one bit is $m_{bit} = k_b T \ln(2)/C^2$, where $k_b \approx 1.38 \times 10^{-23}$ j/kg is Boltzmann's constant, $T$ is the absolute temperature of the information carrier, and $C \approx 3.0 \times 10^8$ m/s is the speed of light in vacuum. Therefore, if the carrier of information $I$ is matter with a mass of 1 kilogram and its energy does not dissipate, the upper bound of the information volume that can be carried is always $\text{volume}_\sigma(I) = 1/m_{bit} = C^2/k_b T \ln(2)$ bits, where the measure $\sigma$ is in bits. When $T$ is at a normal temperature of 300 $K^o$, $\text{volume}_\sigma(I)$ is approximately $10^{37}$ bits. Because this formula is derived from principles of classic thermodynamics, it is applicable to only the equilibrium state of classic digital memory and cannot be applied to the case of quantum carriers such as electrons and photons.

For information with a quantum as its carrier, formula (2) provides the information volume that a single quantum carrier can hold at a specific time. Due to the disjoint property of quantum carriers, the additivity of information volume shown in formula (1) can help us to estimate the information volume that a quantum carrier can hold.

**Theorem 1** (The relationship formula of information, matter, energy, and time) Assume that the carrier $c$ of restorable information $I = \langle o, T_h, f, c, T_m, g \rangle$ is composed of $N$ quanta and that one part of carrier $c$ exists in the form of matter and the other part exists in the form of radiation, with mass $m$ and energy $E_r$, respectively. Assume further that $E$ is the total energy of $c$, measure $\sigma$ is the number of all states of $c$ over the reflection time duration $T_m$, and $t$ is the length of the reflection time duration of information. Then, when $t$ is sufficiently large, $\text{volume}_\sigma(I) = 4Et/h = 4(mC^2 + E_r)t/h$ (qubit). When $t$ is or approaches zero, $\text{volume}_\sigma(I) = N$. To express the formula in simpler form, we let $I$ directly represent its information volume. Then, a very concise formula of information volume can be obtained:

$$I = \begin{cases} 4Et/h = 4(mC^2 + E_r)t/h, & t \text{ is sufficiently large} \\ N, & t \text{ is or is approaching } 0, \end{cases} \quad (3)$$

where $I$ has the unit of qubits, $C$ is the speed of light and $h$ is the Planck constant.

**Proof** Assume the total energy of carrier $c$ is $E$ for restorable information $I = \langle o, T_h, f, c, T_m, g \rangle$. This total energy remains a constant $E$ at any moment in the reflection time duration $T_m$ according to the basic principle of the conservation of energy [38]. In addition, carrier $c$ is composed of $N$ quanta. If each individual quantum has average energy $\Delta E$, then the number of quanta is $N = E/\Delta E$ within the reflection time duration $T_m$. Without loss of generality, we can assume that $c = \bigcup_{i=1}^{N} c_i$, where each $c_i (i = 1,,, N)$ is a single quantum that serves as a carrier in the reflection time duration $T_m$. Let $I_i$ denote the subinformation carried by $c_i$. Each $I_i$ is a single quantum carrier's information, and $t$ is the length of the reflection time duration of information $I$ and subinformation $I_i$. From equation (2), we have $\text{volume}_\sigma(I_i) = 4\Delta Et/h$ (when $t$ is sufficiently large) or $\text{volume}_\sigma(I_i) = 1$ (when $t$ is or approaches zero), $(i = 1,,, N)$. Since we also have $I = \bigcup_{i=1}^{N} I_i$, each $c_i$ is a distinct quantum, this subinformation does not overlap, and $I_i$ is a combination of $[4\Delta Et/h] + 1$ atomic information. By the additivity of measure $\sigma$ over mutually disjoint sets and formula (3), we have

$$\text{volume}_\sigma(I) = \sum_{i=1}^{N} \text{volume}_\sigma(I_i) \begin{cases} \approx \sum_{i=1}^{E/\Delta E} 4\Delta Et/h = 4Et/h (\text{qubit}), & t \text{ is sufficiently large} \\ = \sum_{i=1}^{N} 1 = N (\text{qubit}), & t \text{ is or is approaching } 0. \end{cases}$$

Furthermore, the total energy $E$ of $c$ is the sum of the energy contained in its matter form and the energy in its radiation form. By the matter-energy conversion formula [49][50], the matter form has energy $mC^2$, where $C$ is the speed of light. Then, $E = mC^2 + E_r$. Thus, when $t$ is sufficiently large, we have $\text{volume}_\sigma(I) = 4(mC^2 + E_r)t/h$(qubit). The theorem is proved.

Since formula (3) covers the basic measurements of information, matter, energy, and time with a simple and clear expression, we call it the relationship formula of information, matter, energy, and time. Imagine that we have an information carrier that can be decomposed into pure quanta [38] and that the energy it possesses is equivalent to 1 kg of matter. By formula (3), the information volume carried over 1 second of time is $4C^2/h \approx 5.3853 \times 10^{50}$ (qubit). Additionally, at any moment, the information volume it can carry is exactly the number of quanta it contains. If this carrier is composed of all electrons, since the mass of a single electron is approximately $9.1 \times 10^{-31}$ kg, i.e., a one-kg carrier has approximately $10^{30}$ electrons, the information volume that can be carried is approximately $10^{30}$ (qubit). Alternatively, we could assume that this carrier is composed of all photons. Since photons are mainly at an energy of 0.2 meV [38], i.e., $0.2 \times 10^{-3}$ eV, which is equivalent to the energy of 1 kilogram of matter and should be $C^2 \approx 9.0 \times 10^{16}$ j $\approx 5.6 \times 10^{35}$ eV and the number of photons it contains is approximately $2.8 \times 10^{39}$, the information volume that the carrier can carry at this time is approximately $10^{39}$ (qubit).

The vast universe is full of all kinds of information and is thought to be a gigantic quantum computer. During past decades, researchers have explored the question of how much information the whole universe may have possibly carried to date [51][52][53][54][55]. The formula for the relationship between information and matter, energy, and time from Theorem 1 and the basic principle of energy conservation enable us to answer this question very easily.

Standard inflationary theory predicts a spatially at universe. Einstein's general relativity determines that such a universe has a total energy density equal to the critical density $\rho_c = 3H_0^2/8\pi G$ [46], where $H_0 \approx 2.1 \times 10^{-18}$ / s is the current Hubble parameter and $G \approx 6.7 \times 10^{-11}$ m³ / kgs² is the gravitational constant, i.e., $\rho_c \approx 7.9 \times 10^{-27}$ kg / m³. On the other hand, the radius of the observable universe is approximately $L = 4.56 \times 10^{10}$ light years [56], from which the volume of the whole universe can be estimated as $V \approx (4/3) \pi L^3 \approx 3.35 \times 10^{80}$ m³. Therefore, the total mass of the universe can be estimated as $m = \rho_c V \approx 2.6 \times 10^{54}$ kg. According to the common understanding, the age of the universe is approximately 13.7 billion years [57], that is, the length of the information reflection time duration is $t \approx 4.3 \times 10^{17}$ s. According to formula (3), the information volume of the universe to date is $I = 4mC^2t/h \approx 4 \times 2.6 \times 10^{54} \times (3.0 \times 10^8)^2 \times 4.3 \times 10^{17}/(6.6 \times 10^{-34}) \approx 6.1 \times 10^{122}$ (qubit). This estimate is almost the same as the number of logical operations of the universe in [52]: $10^{123}$. According to OIT, any logical operation of the universe must create a particular state, and these states must contain objective information. Therefore, the number of logical operations of the universe at a particular time is the same as the number of states the universe has at that time. On the other hand, this estimate is not the same as the $10^{90}$ bits of information capacity of the universe estimated in [52]. This is because the information capacity is defined through entropy [58], which differs greatly from the definition of information volume in Theorem 1. By comparison, we believe that it is more universally applicable to define the information volume as "the measure of all states of the carrier over the reflection time". More importantly, Theorem 1 establishes the general relationships among the three major elements, namely, matter, energy, and information of the objective world, and time. For any information carrier, if the amount of mass, energy, and time are known, we can apply formula (3) to obtain the upper bound of the information volume.

## 5 Concluding remarks

Based on the basic concept of OIT, we propose four postulates on information and construct a sextuple model that satisfies these postulates as necessary and sufficient conditions. We also unify the relevant classical and commonly used information principles through eleven types of information metrics defined by OIT, which shows that the theoretical system of OIT has excellent potential for broader interpretation and application. We determine the information volume that a single quantum carrier can carry over its reflection time duration and establish a formula for the relationship between information and matter, energy, and time, revealing a deep and more precise relationship among the basic elements of the objective world. This study further highlights the scientific significance in exploring and forming a universal consensus on information, establishing a comprehensive and universal information measurement system, and providing a theoretical reference for the research and application of information science.

OIT defines eleven types of information metrics and analyzes the attributes of information, including its objectivity, restorability, combination, transitivity, and correlation. In this paper, we study the essence of information and its relationship with matter, energy and time and the relationship between information metrics and some classical information principles. Further in-depth research could address the properties and relationships among these components with regard to other attributes and measurements of information. Moreover, by comparing and testing the compatibility with other existing principles and algorithms of information science, we can help to build a more complete and applicable theoretical system of information science and a more comprehensive ISD to provide a stronger theoretical instrument for the development and application of information technology.

**Acknowledgment**

Professors Zhenyu Liu and Yashi Wang from China University of Political Science and Law, Professors Shuliang Wang and Tao Zheng, Mr. Yingxu Dang from Beijing Institute of Technology, and Dr. Yingfei Wang from the Information Technology Service Center of the Supreme People's Court have offered indispensable assistance to the author with regard to research ideas and literature collection during joint study on OIT and information system dynamics. Mr. Ding Ding from People's Court Press assisted with all figures. Sincere appreciation is expressed for all their help and assistance.

This study is supported by the National Key R&D Program of China (2016YFC0800801).

**Data availability**

The data that support the findings of this study are available from the corresponding author on reasonable request and all other data generated or analyzed during the study are included in the present manuscript.

**References**

[1] Shannon, C. E. The mathematical theory of communication. The Bell System Technical Journal. 27, 379-423, 623-656 (1948).

[2] Ryali, C. K. et al. From likely to likable: The role of statistical typicality in human social assessment of faces. Proceedings of the National Academy of Sciences of the United States of America 117 (47) , 29371-29380 (2021)

[3] Skolnik M.I. Radar handbook. 2nd ed. New York: McGraw-Hill(1990).

[4] Ulaby, F. T., Dobson, M. C. The Radar Equation, Handbook of Radar Scattering Statistics for Terrain. Artech House, 685 Canton ST, Norwood, MA 02062 USA (2019).

[5] Nyquist H. Certain topics in telegraph transmission theory. Proc IEEE, 90(2):280–305(1928).

[6] Holme, H. C. M. et al. ENLIVE: An Efficient Nonlinear Method for Calibrationless and Robust

Parallel Imaging. Scientific Reports 9, 3034 (2019).

[7] Shapiro C, Varian H.R. Information rules: a strategic guide to the network economy. Boston: Harvard Business School Press(1998).

[8] Li, Y. P. et al. Optimal Pricing for Peer-to-Peer Sharing With Network Externalities. IEEE-ACM Transactions on Networking 29(1), 148-161 (2021).

[9] Kalman, R. E. A new approach to linear filtering and prediction problems. J Basic Eng , 82(1):35–45(1960).

[10] Stojic, H. et al. Uncertainty in learning, choice, and visual fixation. Proceedings of the National Academy of Sciences of the United States of America 117 (6) , 3291-3300 (2020).

[11] Hartley, R. V. L. Transmission of Information. The Bell System Technical Journal VII, 535 - 563 (1928).

[12] Brillouin, L. Science and Information Theory, Academic Press Inc., New York (1956).

[13] MacKay, D. Information, Mechanism and Meaning; MIT Press: Cambridge, MA, USA (1969).

[14] Campbell, J. Grammatical Man: Information, Entropy, Language, and Life; Simon and Schuster: New York, NY, USA, 32(1982).

[15] Landauer, R. Information is physical. Phys. Today 44, 23-29 (1991).

[16] Fleissner P, Hofkirchner W. Emergent information: towards a unified information theory. BioSyst, 38: 243-248(1996).

[17] Zhong, Y. X. Principles of Information Science. 3rd ed. Beijing: Beijing University of Posts and Telecommunications Press, 111-116(2002).

[18] Rao M, Chen Y, Vemuri B C, et al. Cumulative residual entropy: a new measure of information. IEEE Trans Inform Theory, 50: 1220-1228(2004).

[19] Zaliwski A S. Information-is it subjective or objective? Triple C: communication, capitalism & critique. Open Access J Global Sust Inform Soc , 9: 77-92 (2011)

[20] Yan, X. S. Information science: Its past, present and future[J]. Information, 2(3): 510-527 (2011).

[21] Burgin M. Theory of Information: Fundamentality, Diversity and Unification. Singapore: World Scientific Publishing Co Pty Ltd., (2010).

[22] Li X, Cheng Y Q, Wang H Q, et al. Progress in theory and applications of information geometry. Sci Sin Inform , 43: 707-732 (2013).

[23] Logan, R. K. What Is Information?: Why Is It Relativistic and What Is Its Relationship to Materiality, Meaning and Organization. Information. 3, 68-91 ( 2012).

[24] Xu, J.f., Tang, J., & Ma, X.F. et al. Research on The Model and Measurement of Objective Information. Science China: Information Science. 45(3): 336-353 (2015).

[25] Xu, J. F., Wang, S.L., Liu, Z.Y., & Wang, Y. F. Objective Information Theory Exemplified in Air Traffic Control System. Chinese Journal of Electronics. 30(4), 743-751(2021).

[26] Xu, J.F. et al. Foundations and applications of information systems dynamics, Engineering, doi: https://doi.org/10.1016/j.eng.2022.04.018 (2022).

[27] Xu, J. F., Sun, F. H., & Chen, Q. W. Introduction to the Smart Court System-of-Systems Engineering Project of China. Singapore: Spinger, (2022).

[28] Wiener, N. Cybernetics: Or Control and Communication in the Animal and the Machine. MIT Press, New York( 1961).

[29] Program on Information Resources Policy. Program Projects. Annual Report Volume 2 (1975-1976). Cambridge(MA): Computation Laboratory, Harvard University, Report No.:R-76-2 (1976).

[30] Landauer, R. The physical nature of information. Phys. Lett. A 217, 188-193 (1996).

[31] Rayleigh L. LVI investigations in optics, with special reference to the spectroscope. Philos Mag Ser 5, 8(51):477-486(1879).

[32] Lienig J, Bruemmer H. Reliability analysis. In: Fundamentals of Electronic Systems Design. Cham: Springer International Publishing, 45–73 (2017).

[33] Flores I, Madpis G. Average binary search length for dense ordered lists. Commun ACM,14(9):602–3(1971).

[34] Zhou, Minqiang. Real Analysis (Second Edition). Peking University Press, Beijing, (2008).

[35] Zhang, X. et al. Semiconductor quantum computation, National Science Review 6(1), 32-54 (2019)

[36] Guerreschi, G. G., Matsuura, A. Y. QAOA for Max-Cut requires hundreds of qubits for quantum speed-up, Scientific Reports 9, 6903 (2019).

[37] Wiener, N. The Human Use of Human Beings: Cybernetics and Society(New Edition). Da Capo Press, Cambridge,(1988).

[38] Taube, M. Evolution of Matter and Energyon a Cosmic and Planetary Scale. Springer-Verlag, New York,(1985).

[39] Zizzi, P. Holography, Quantum Geometry, and Quantum Information Theory. Entropy 2, 36-39 (2000).

[40] Lloyd, S. Ultimate physical limits to computation. Nature 406, 1047-1054 (2000).

[41] Margolus, N. & Levitin, L. B. The maximum speed of dynamical evolution. Physica D 120, 188-195 (1998).

[42] Nielsen, M. A. & Chuang, I. L. Quantum Computation and Quantum Information. Cambridge University Press, New York, (2010).

[43] Li, Z. K. et al. Resonant quantum Principal component analysis. Sci. Adv. 7, eabg2589, Doi: 10.1126/sciadv.abg 2589 (2021).

[44] Smith, A. W. R. et al. Qubit readout error mitigation with bit-flip averaging. Sci. Adv. 7, eabi8009, Doi: 10.1126/sciadv.abi 8009 (2021).

[43] Li, Z. K. et al. Resonant quantum Principal component analysis. Sci. Adv. 7, eabg2589, Doi: 10.1126/sciadv.abg 2589 (2021).

[44] [44] Smith, A. W. R. et al. Qubit readout error mitigation with bit-flip averaging. Sci. Adv. 7, eabi8009 (2021), Doi: 10.1126/sciadv.abi 8009.

[45] Hubble, E. Proc. Natl. Acad. Sci. U.S.A. 15, 168 (1929).

[46] Bahcall, N. A. & Ostriker, J. P. & Perlmutter, S. & Steinhardt, P. J. The Cosmic Triangle: Revealing the State of the Universe. Gen. Science 284,1481–1488 (1999).

[47] Landauer, R. Irreversibility and heat generation in the computing process. IBM Journal of Research and Development 5(3), 183 - 191 (1961).

[48] Vopson, M. M. The mass-energy-information equivalence principle. AIP Advances 9, 095206:1-4(2019).

[49] Einstein, A. On The Electrodynamics of Moving Bodies. Annalen der Physik. 17(10), 891-921 (1905).

[50] Rohrlich, F. An elementray derivation of E = mc2. Am.J.Phys.58(4),348-349 （1990）.

[51] Gleick, J. The Information: A History, a Theory, a Flood. (Vintage, New York, 2011).

[52] Lloyd, S. Computational Capacity of the Universe. Physical Review Letters, 88(23), 237901:1-4 (2002).

[53] Vopson, M.M. Estimation of the information contained in the visible matter of the universe.

AIP Advances 11,105317:1-5 (2021).

[54] Davies, P.C.W. "Why is the physical world so comprehensible?" in Complexity,Entropy and the Physics of Information.  Addison-Wesley,Redwood City (1990).

[55] Treumann, R.A. Evolution of the information in the universe. Astrophys.Space Sci. 201, 135-147 (1993).

[56] Gott III, J. R., Jurić, M., Schlegel, D., Hoyle, F. , Vogeley, M., Tegmark,M., Bahcall, N. & Brinkmann, J. A map of the universe. Astrophys.J.624(2),463-484(2005).

[57] Bryson, B. A Short History of Nearly Everything. Broadway Books, New York (2003).

[58] Reichl, L. E. A Modern Course in Statistical Physics. University of Texas Press, (1980).